\documentclass[conference]{IEEEtran}
\IEEEoverridecommandlockouts

\usepackage{cite}
\usepackage{amsmath,amssymb,amsthm,amsfonts,multicol, nccmath}
\usepackage{algorithm, multirow, ragged2e}
\usepackage[noend]{algpseudocode}
\usepackage{graphicx, lipsum}
\usepackage{xcolor}
\usepackage{tikz}
\usepackage{pgfplots}
\pgfplotsset{compat=1.18}
\usetikzlibrary{patterns}
\usepackage{balance}
\usepackage[hidelinks, draft]{hyperref}
\usepackage{subcaption}

\usepackage{float, array, tabularray, diagbox}
\UseTblrLibrary{diagbox}
\def\BibTeX{{\rm B\kern-.05em{\sc i\kern-.025em b}\kern-.08em
		T\kern-.1667em\lower.7ex\hbox{E}\kern-.125emX}}

\begin{document}

\title{Vision-Based Learning for Cyberattack Detection in Blockchain Smart Contracts and Transactions}

\author{\IEEEauthorblockN{Do Hai Son\IEEEauthorrefmark{1}\IEEEauthorrefmark{5}, Le Vu Hieu\IEEEauthorrefmark{2}, Tran Viet Khoa\IEEEauthorrefmark{3}, Yibeltal F.  Alem\IEEEauthorrefmark{3},\\Hoang Trong Minh\IEEEauthorrefmark{4}, Tran Thi Thuy Quynh\IEEEauthorrefmark{2}, Nguyen Viet Ha\IEEEauthorrefmark{2}, and Nguyen Linh Trung\IEEEauthorrefmark{2}.
    \vspace{0.2cm}\\}
    \IEEEauthorrefmark{1} VNU Information Technology Institute, Hanoi, Vietnam. \\
    \IEEEauthorrefmark{2} VNU University of Engineering and Technology, Hanoi, Vietnam. \\
    \IEEEauthorrefmark{5} School of Electrical Engineering, Computing and Mathematical Sciences, Curtin University, Australia. \\
    \IEEEauthorrefmark{3} University of Canberra, Australia. \\
    \IEEEauthorrefmark{4} Posts and Telecommunications Institute of Technology, Vietnam.
    \thanks{Corresponding author: Tran Viet Khoa (khoa.tran@canberra.edu.au).}
}

\maketitle

\begin{abstract}
Blockchain technology has experienced rapid growth and has been widely adopted across various sectors, including healthcare, finance, and energy. However, blockchain platforms remain vulnerable to a broad range of cyberattacks, particularly those aimed at exploiting transactions and smart contracts (SCs) to steal digital assets or compromise system integrity. To address this issue, we propose a novel and effective framework for detecting cyberattacks within blockchain systems. Our framework begins with a preprocessing tool that uses Natural Language Processing (NLP) techniques to transform key features of blockchain transactions into image representations. These images are then analyzed through vision-based analysis using Vision Transformers (ViT), a recent advancement in computer vision known for its superior ability to capture complex patterns and semantic relationships. By integrating NLP-based preprocessing with vision-based learning, our framework can detect a wide variety of attack types. Experimental evaluations on benchmark datasets demonstrate that our approach significantly outperforms existing state-of-the-art methods in terms of both accuracy (achieving 99.5\%) and robustness in cyberattack detection for blockchain transactions and SCs.
\end{abstract}


\section{Introduction}\label{Intro}
Blockchain technology has advanced rapidly in recent years, particularly in the area of data management. By storing data on a distributed ledger, blockchain provides key properties such as immutability, security, and transparency. These properties ensure that data remains tamper-resistant, securely stored, and transparently accessible throughout the network, which strengthens the overall integrity and trustworthiness of the system. Its core features, including decentralization, immutability, and fault tolerance, make it a promising solution for a wide range of applications. 
As a result, blockchain has been increasingly adopted in sectors such as finance, healthcare, energy, and the Internet-of-Things (IoT)~\cite{yue2021survey}. Despite these advantages, many blockchain systems have recently become targets of security threats, particularly in transactions and SCs. For instance, as of December 2024, data from the DeFiHacksLabs Reproduce Repository reports over 150 smart contract attack incidents in 2024 alone, resulting in losses exceeding \$328 million~\cite{Defihack, Khoa2024WCNC}. Although most attacks so far have focused on financial systems, the increasing use of blockchain in critical sectors, such as healthcare and energy, raises concerns that future attacks may have direct and serious consequences for human well-being~\cite{chen2020survey}.

Several studies have explored cyberattack detection in blockchain SCs and transactions (txs). In~\cite{wang2020contractward}, the authors proposed ContractWard, a system that applied machine learning models to detect six types of attacks, achieving an F1-score of up to 97\%. However, their approach analyzed the source code of SCs rather than the bytecode. This presents a limitation, as only bytecode is available once a contract is deployed and executed through transactions. Therefore, direct analysis of bytecode is crucial for real-time attack detection. In~\cite{nguyen2019detect}, the authors used predefined attack vectors to analyze bytecode directly. While effective for specific attacks, their method was evaluated on a small dataset of around 100 samples, limiting its ability to detect new or evolving threats. In~\cite{khoa2023collaborative}, the authors proposed a machine learning (ML) framework that performed real-time detection of six types of attacks by directly analyzing bytecode in SCs and transactions. They also introduced the BTAT dataset, which includes labeled cyberattack samples from blockchain environments. Experimental results showed their framework achieved an accuracy of up to 94\% on this dataset. In~\cite{2024SonISCIT}, the authors developed an anomaly detection model for blockchain-based supply chain systems using only network-layer traffic data. While their approach achieved a promising accuracy of 96.5\%, it was limited to general anomaly detection and did not target specific cyberattack types within SCs or transaction-level behaviors.

There are several challenges in detecting cyberattacks in SCs and transactions within blockchain systems. Firstly, improving the accuracy of cyberattack detection is crucial. Low accuracy can lead to missed attacks or false alarms, resulting in financial losses, security breaches, and reduced trust in blockchain systems. In contrast, high accuracy enhances detection reliability and supports stronger protection for decentralized applications. Secondly, when an SC is deployed, it is converted into bytecode, which is a sequence of hexadecimal values that represents the core logic of the contract and its associated transactions. Analyzing this bytecode is essential for real-time attack detection~\cite{huang2021hunting}. There are two common approaches for bytecode analysis: using the source code or analyzing the bytecode directly. However, only about 1\% of SCs' source code is publicly available, making source-based analysis infeasible in most cases, while direct bytecode analysis is often time-consuming and less reliable~\cite{huang2021hunting}.

To address the first challenge, we propose a novel framework for detecting cyberattacks in SCs and transactions with higher accuracy than existing methods. Our approach uses a ViT-based architecture, which allows the model to capture complex patterns more effectively than traditional methods. This makes our solution not only more accurate but also more flexible and easier to extend for detecting additional types of attacks compared to existing vector-based approaches. To tackle the second challenge, we design a preprocessing framework that uses NLP techniques to convert blockchain transaction features, such as opcodes (a structured representation of bytecode), into image representations. In addition, we extract and analyze critical transaction attributes, including transaction value, gas usage, and input length, to enrich the feature set and improve detection accuracy. We carefully study the importance of each feature and apply specific preprocessing approaches to improve its usefulness for attack detection. 
Experimental evaluations on the BTAT~\cite{khoa2023collaborative} dataset demonstrate that our model achieves a classification accuracy of 99.5\% compared to baseline ML models (LR, KNN, SVM) and deep learning (DL) models (i.e., CNN, ResNet, MobileNetv2). Remarkably, our proposed ViT-based model uses only 8\% of the trainable parameters of ResNet~\cite{he2016deep}, indicating a lightweight architecture and improved computational efficiency without loss of~accuracy.

\section{Our Proposed Framework}
This section introduces the key components of our proposed framework for cyberattack detection in blockchain SCs and transactions. First, we present the system model to clarify the deployment context and operational assumptions. Next, we explain the NLP-based preprocessing technique designed to convert transaction data into a unified image representation. Then, we describe the vision-based learning architecture employed for processing and classification.

\subsection{System model}\label{sec:model}
\begin{figure}
    \centering
    \includegraphics[width=\linewidth]{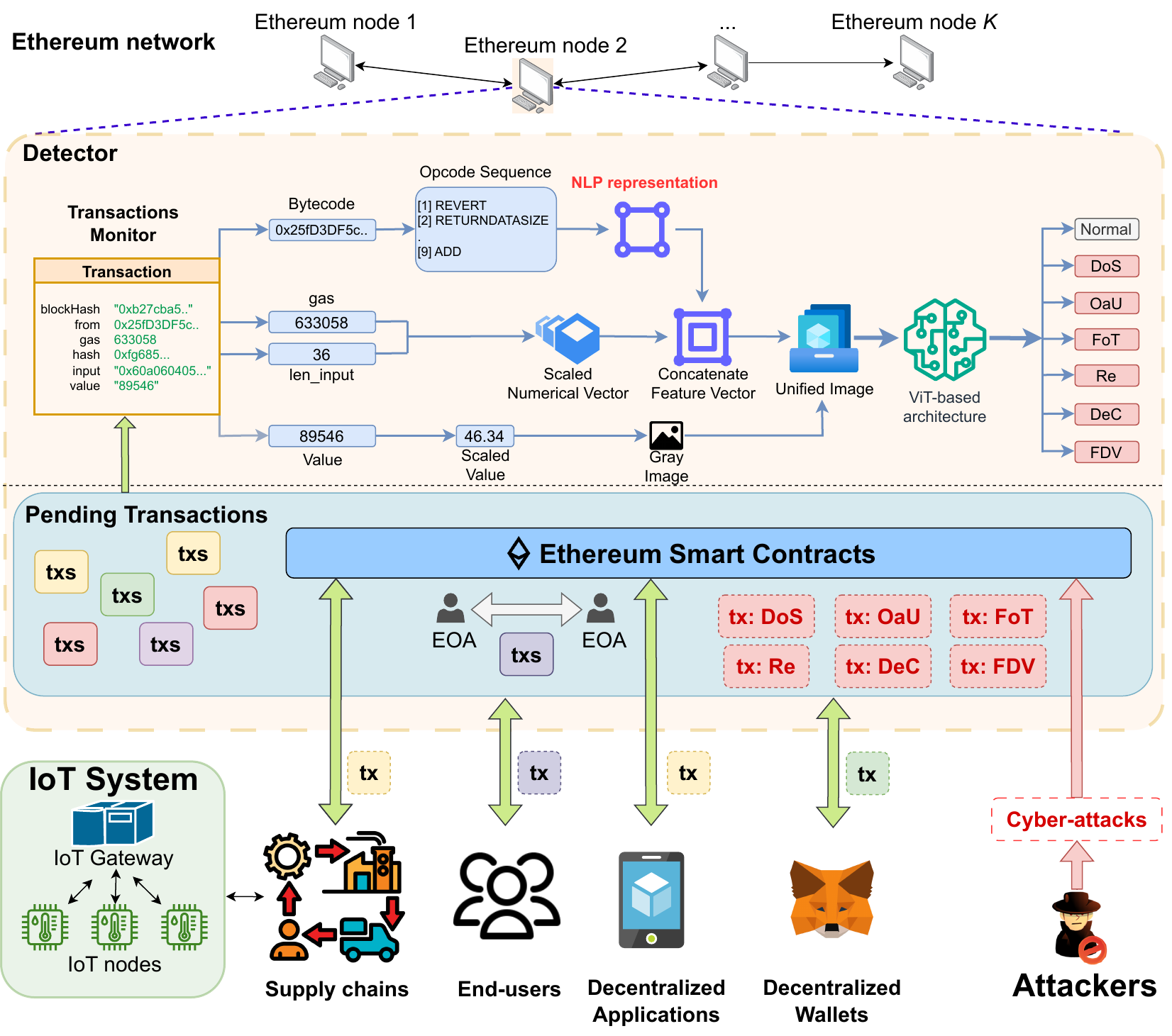}
    \caption{The proposed model of cyberattack detection in Ethereum smart contracts and transactions.}
    \label{fig:system_model}
\end{figure}
In this work, we focus on cyberattack detection in Ethereum SCs and transactions. As shown in Fig.~\ref{fig:system_model}, we assume that transactions at each Ethereum node (a.k.a, end-point) originate from various agencies, each with distinct purposes. For example, end-users exchange the native Ethereum token (ETH) either directly or through decentralized wallets such as MetaMask. Decentralized applications (e.g., mobile games, IoT, and supply chain platforms) also rely on Ethereum SCs to store digital assets and information. This information is considered trustworthy due to the immutability of blockchain technology. However, in the same manner, attackers can exploit intrinsic vulnerabilities in the Ethereum protocol and the consensus mechanism, as well as weaknesses in SCs, to disrupt the network or steal digital assets and valuable information~\cite{chen2020survey}.

To detect these malicious SCs and transactions, we deploy a detector at the top of each Ethereum node, as illustrated in Fig.~\ref{fig:system_model}. Therefore, this detector can collect pending transactions (e.g., transactions and SCs' deployments) from the Ethereum node. The process of classifying a transaction consists of three stages. First, relevant fields inside a transaction, i.e., input (a.k.a., bytecode), length of input (len\_input), gas, and value, are extracted. Second, these fields are processed to generate an NLP representation for the input, and normalized into vectors for gas, len\_input, and value. The outputs of these processes are then fused into a unified image representation applicable to both SCs and transactions. Finally, our proposed DL model, based on the ViT architecture, takes this image as input and outputs a classification result to alert the administrator/end-users if malicious behavior is detected.

\subsection{NLP-based Preprocessing Technique}
\begin{figure*}[t!]
    \centering
	\includegraphics[width=\linewidth]{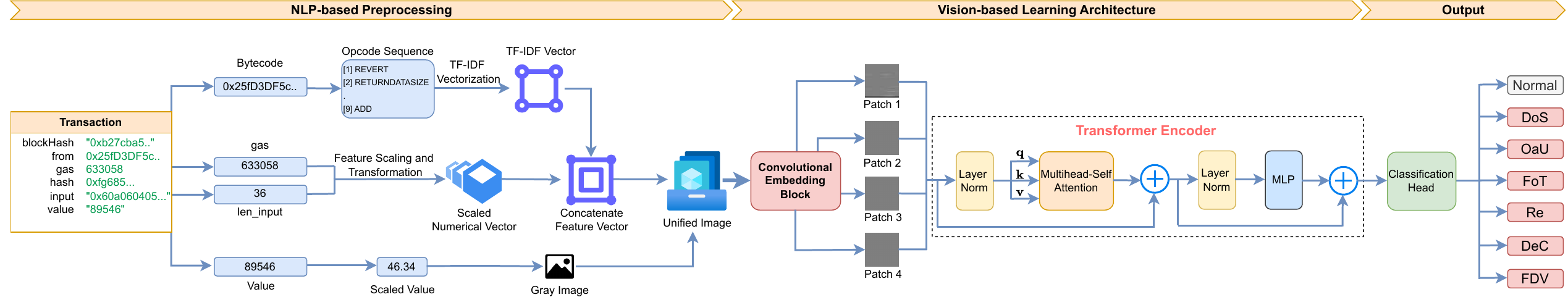}
	\caption{The proposed NLP-based preprocessing and ViT-based architecture.}
	\label{fig:Processing}
\end{figure*}

During preprocessing, each transaction $\mathcal{T}_i \in \boldsymbol{\mathcal{T}}$, where $\boldsymbol{\mathcal{T}}$ denotes the dataset, is transformed into an image $\mathbf{I}_i$. This process consists of three main steps: feature extraction, feature assembly, and image construction.

\subsubsection{Opcode-based Feature Extraction via TF-IDF}

This stage captures the semantic content of transaction bytecode. Each transaction $\mathcal{T}_i$ consists of several fields, e.g., the input of hexadecimal bytecode $b_i$, gas usage $g_i$, value $v_i$, and so on. Specifically, the input length $\ell_i$ is computed as the number of bytes in $b_i \in \mathcal{T}_i$. To begin, we focus on the opcode-level semantics derived from the bytecode. The bytecode $b_i$ is decoded into an opcode sequence $s_i$ using a disassembly function $\Phi_{\text{op}}$:
\begin{equation}
s_i = \Phi_{\text{op}}(b_i).
\end{equation}

The obtained opcode sequence $s_i$ is treated as a text document, following an NLP perspective. This allows statistical representation through the Term Frequency–Inverse Document Frequency (TF-IDF) method~\cite{manning2008}. For each term (i.e., opcode) $t$ in $s_i$, the term frequency (TF) is given by:
\begin{equation}
\text{TF}(t, s_i) = f_{t,s_i},
\end{equation}
where $f_{t,s_i}$ denotes the raw count of term $t$ in document $s_i$. The inverse document frequency (IDF) is calculated with respect to the corpus $\mathcal{S}_{\text{train}}$, i.e., the set of opcode sequences extracted from the training transactions, as follows~\cite{manning2008}:
\begin{equation}
\text{IDF}(t, \mathcal{S}_{\text{train}}) = \log{\frac{|\mathcal{S}_{\text{train}}|}{|\{s \in \mathcal{S}_{\text{train}}: t \in s\}| + 1}}.
\end{equation}

The final TF-IDF score is the product of the two components~\cite{manning2008}:
\begin{equation}
\text{TF-IDF}(t, s_i, \mathcal{S}_{\text{train}}) = \text{TF}(t, s_i) \cdot \text{IDF}(t, \mathcal{S}_{\text{train}}).
\end{equation}

Let $\mathcal{V}_{\text{op}} = \{t_1, t_2, \ldots, t_{d_{\mathrm{op}}}\}$ be the vocabulary of all unique opcodes extracted from $\mathcal{S}_{\text{train}}$. Applying the TF-IDF procedure to all opcodes in the vocabulary yields a fixed-dimensional feature vector:
\begin{equation}
\begin{aligned}
    \mathbf{u}_{\mathrm{op},i} = [&\text{TF-IDF}(t_1, s_i, \mathcal{S}_{\text{train}}), \text{TF-IDF}(t_2, s_i, \mathcal{S}_{\text{train}}), \\&\ldots, \text{TF-IDF}(t_{d_{\mathrm{op}}}, s_i, \mathcal{S}_{\text{train}})]^T,
\end{aligned}
\end{equation}
where $\mathbf{u}_{\mathrm{op},i} \in \mathbb{R}^{d_{\mathrm{op}}}$ represents the semantic structure of transaction $\mathcal{T}_i$, and $d_{\mathrm{op}}$ is the size of the opcode vocabulary.

\subsubsection{Feature Assembly and Image Transformation}

In parallel, scalar attributes such as the gas usage $g_i$ and bytecode length $\ell_i$ are normalized using standard-scaler to form the numerical attribute vector, as follows:
\begin{equation}
\mathbf{m}_i = [g_i, \quad \ell_i],
\end{equation}
\begin{equation}
\mathbf{u}_i = \frac{\mathbf{m}_i - {\mu}_m}{{\sigma}_m},
\end{equation}
where $\mathbf{u}_i \in \mathbb{R}^{d_u}$ with $d_u$ is the number of numerical features; ${\mu}_\mathbf{m}$ and ${\sigma}_\mathbf{m}$ are the mean and standard deviation of $\mathbf{m}$ computed on $\boldsymbol{\mathcal{T}}$, respectively. The final composite attribute vector is obtained via the following horizontal concatenation:
\begin{equation}
\mathbf{x}_i = [\mathbf{u}_i, \quad \mathbf{u}_{\mathrm{op},i}],
\end{equation}
with $\mathbf{x}_i \in \mathbb{R}^{d_u + d_{\mathrm{op}}}$. This vector $\mathbf{x}_i$ is then interpolated and reshaped into a gray-scale matrix (gray image) $\mathbf{I}'_i \in \mathbb{R}^{H' \times W}$ through bilinear interpolation from $\mathbb{R}^{d_u + d_{\mathrm{op}}} \rightarrow \mathbb{R}^{H' \times W}$, given~by:
\begin{equation}
I'_{i,h,w} = \mathbf{r}_i(hW + w + 1),
\end{equation}
where $\mathbf{r}_i \in \mathbb{R}^{H' \times W}$ is the bilinearly interpolated vector obtained by resizing $\mathbf{x}_i$, and $h \in \{0, 1, \ldots, H'-1\}$, $w \in \{0, 1, \ldots, W-1\}$. The value field is normalized as $a_i = 255\left( \log_{10}(v_i) / 18 \right)$. The normalized value $a_i$ is then broadcast into a row vector $\mathbf{a}_i \in \mathbb{R}^{1 \times W}$ and appended as the last row of the image, as follows:
\begin{equation}
\mathbf{I}_i = [\mathbf{I}'_i, \quad \mathbf{a}_i]^T.
\end{equation}

This image $\mathbf{I}_i \in \mathbb{R}^{H \times W}$ serves as the input for our vision-based learning model.

\subsection{Vision-based Learning Architecture}

In this work, we propose a ViT-based architecture that combines Vision Transformers~\cite{dosovitskiy2020image} for detecting cyberattacks from preprocessed images with a hierarchical feature extraction mechanism inspired by ResNet~\cite{he2016deep}. To enable this, we use a convolutional embedding block that performs multi-stage convolutional patch embedding.
Each input image is represented as $\mathbf{I}_i \in \mathbb{R}^{H \times W}$ and is first expanded into a tensor with a single channel. It is then passed through three sequential convolutional layers to extract feature representations~\cite{he2016deep}:
\begin{equation}
\mathbf{P}_i = \text{Conv}_3(\text{Conv}_2(\text{Conv}_1(\mathbf{I}_i))),
\end{equation}
where $\text{Conv}_1: \mathbb{R}^{1 \times H \times W} \rightarrow \mathbb{R}^{\frac{D}{4} \times H \times W}$,
$\text{Conv}_2: \mathbb{R}^{\frac{D}{4} \times H \times W} \rightarrow \mathbb{R}^{\frac{D}{2} \times \frac{H}{2} \times \frac{W}{2}}$, and
$\text{Conv}_3: \mathbb{R}^{\frac{D}{2} \times \frac{H}{2} \times \frac{W}{2}} \rightarrow \mathbb{R}^{D \times \frac{H}{4} \times \frac{W}{4}}$ with $D$ as the embedding dimension of the Transformer. The resulting tensor $\mathbf{P}_i$ is flattened into a sequence of patches and concatenated with a learnable class token $\mathbf{z}_{\text{cls}} \in \mathbb{R}^{1 \times D}$ and positional embeddings $\mathbf{E}_{\text{pos}} \in \mathbb{R}^{(N+1) \times D}$, where $N = \frac{HW}{16}$~\cite{dosovitskiy2020image}:
\begin{equation}
\mathbf{Z}^0_i = [\mathbf{z}_{\text{cls}}, \quad \text{Flatten}(\mathbf{P}_i)] + \mathbf{E}_{\text{pos}} \in \mathbb{R}^{(N+1) \times D}.
\end{equation}

This sequence is then passed through $K$ Transformer encoder blocks using pre-normalization for improved training stability. For each encoder layer indexed by $k \in \{1, \dots, K\}$, LayerNorm is applied before each sub-layer. The output of the multi-head self-attention (MSA) sub-layer is computed as~\cite{dosovitskiy2020image}:

\begin{equation}
\mathbf{Z}^{''}_{i,k} = \text{MSA}(\text{LN}(\mathbf{Z}_{i,k-1})) + \mathbf{Z}_{i,k-1},
\end{equation}
where $\mathbf{Z}_{i,k-1}$ is the output of the previous layer. Each MSA block uses multiple attention heads $l \in \{1, \dots, L\}$ with query, key, and value matrices computed as:
$\mathbf{q}_l = \text{LN}(\mathbf{Z}_{i,k-1}) \mathbf{W}^\mathbf{q}_l$,
$\mathbf{k}_l = \text{LN}(\mathbf{Z}_{i,k-1}) \mathbf{W}^\mathbf{k}_l$, and
$\mathbf{v}_l = \text{LN}(\mathbf{Z}_{i,k-1}) \mathbf{W}^\mathbf{v}_l$,
where $\mathbf{W}^\mathbf{q}_l, \mathbf{W}^\mathbf{k}_l, \mathbf{W}^\mathbf{v}_l \in \mathbb{R}^{D \times \frac{D}{L}}$ are the weight matrices of $\mathbf{q}_l$, $\mathbf{k}_l$, $\mathbf{v}_l$, respectively. With $D_l = \frac{D}{L}$, the self-attention output of each head is~\cite{dosovitskiy2020image}:
\begin{equation}
\mathbf{S}_{i,l} = \mathbf{v}_l \cdot \text{softmax} \left( \frac{\mathbf{q}_l \mathbf{k}^T_l}{\sqrt{D_l}} \right) \in \mathbb{R}^{(N+1) \times \frac{D}{L}}.
\end{equation}

The multi-head attention output is then combined and linearly projected~\cite{dosovitskiy2020image}:

\begin{equation}
\text{MSA}(\text{LN}(\mathbf{Z}_{i,k-1})) = [\mathbf{S}_{i,1}, \mathbf{S}_{i,2}, \ldots, \mathbf{S}_{i,L}] \mathbf{W}_{\text{msa}} \in \mathbb{R}^{(N+1) \times D},
\end{equation}
where $\mathbf{W}_{\text{msa}} \in \mathbb{R}^{D \times D}$ is the output projection matrix of the multi-head self-attention (MSA) module. The output is then passed through a Multilayer Perceptron (MLP) network using pre-normalization~\cite{dosovitskiy2020image}:
\begin{equation}
\mathbf{Z}_{i,k} = \text{MLP}(\text{LN}(\mathbf{Z}^{''}_{i,k})) + \mathbf{Z}^{''}_{i,k},
\end{equation}
where the MLP consists of two linear layers with weights $\mathbf{W}_1 \in \mathbb{R}^{D \times 4D}$ and $\mathbf{W}_2 \in \mathbb{R}^{4D \times D}$.

After $K$ layers, the final representation $\mathbf{Z}_{i,K}$ is used for classification via the class token at index 0~\cite{dosovitskiy2020image}:
\begin{equation}
\mathbf{Y}_i = \mathbf{W}_{\text{head}} \cdot \text{LN}(\mathbf{Z}_{i,K}[0]) \in \mathbb{R}^{C_{\text{out}}},
\end{equation}
where $\mathbf{W}_{\text{head}} \in \mathbb{R}^{D \times C_{\text{out}}}$ is the weight matrix of the classification head, $C_{\text{out}}$ is the number of output classes, and the output $\mathbf{Y}_i$ classifies the transaction as normal or a specific type of~cyberattack.

\section{Experimental Results}\label{Results}
\subsection{Dataset and Evaluation Methods}

To evaluate the performance of our proposed framework, we use the Blockchain Transaction-based Attacks (BTAT) dataset, first introduced in~\cite{khoa2023collaborative}. This dataset was synthesized in a laboratory environment using a private Ethereum network, ensuring that transactions are accurately labeled as either normal or a specific type of attack. The dataset is designed to replicate various real-world attacks that have previously caused significant damage to blockchain systems.

The BTAT dataset includes a wide range of cyberattacks targeting SCs and transactions. The dataset comprises 302,749 transactions in total, including normal (152,423 samples) and six types of attacks as follows:
\begin{itemize}
    \item \textbf{Denial of Service (DoS) with Block Gas Limit:} Attackers exploit functions whose gas requirements can be manipulated to exceed the block gas limit, rendering the contract temporarily unusable (22,994 samples).

    \item \textbf{Overflows and Underflows (OaU):} Arises from integer arithmetic vulnerabilities in Solidity, where variables wrap around upon reaching their maximum or minimum values, enabling attackers to bypass balance checks (29,254 samples).

    \item \textbf{Flooding of Transactions (FoT):} Involves spamming the network with a massive number of meaningless transactions to delay the confirmation of legitimate ones (41,732 samples).
    
    \item \textbf{Re-entrancy (Re):} Exploits vulnerabilities where a contract's state is not updated before an external call, allowing attackers to recursively withdraw funds, as famously occurred in The DAO attack (22,682 samples).
    
    \item \textbf{Delegatecall (DeC):} Involves manipulating the \texttt{delegatecall} mechanism to execute malicious code from an untrusted contract within the context of the main contract, replicating the Parity Multi-Sig Wallet attack (22,455 samples).
    
    \item \textbf{Function Default Visibility (FDV):} Occurs when functions without explicitly defined visibility default to public, allowing unauthorized users to call critical administrative functions, as seen in the first Parity Multi-Sig Wallet hack (11,209 samples).
\end{itemize}

To evaluate the performance of our models, we use several standard metrics that are widely used to evaluate the performance of DL~\cite{confusion_matrix2} and cyberattack detection systems~\cite{khoa2024collaborative}, such as: accuracy, precision, and recall.  
To ensure a fair evaluation across the imbalanced classes, we compute precision and recall using the `macro' averaging method, which calculates the metric independently for each class and then takes the unweighted average.

\subsection{Performance Evaluation and Analysis}
\begin{table}[t]
\centering
\caption{Performances of CNN and ViT-based architectures using baseline and our proposed preprocessing techniques.}
\label{tab:performance_compare}
\begin{tblr}{
  width = \linewidth,
  colspec = {Q[190]Q[177]Q[196]Q[180]Q[190]},
  column{2-5} = {c},
  hlines,
  vlines,
}
 & \textbf{CNN (Baseline)} & \textbf{CNN (Proposed)} & \textbf{ViT-based (Baseline)} & \textbf{ViT-based (Proposed)}\\
\textbf{Accuracy} & 93.8490 & \textbf{98.8582} & 95.1665 & \textbf{99.5200}\\
\textbf{Precision} & 90.4130 & \textbf{98.5740} & 92.0221 & \textbf{99.4393}\\
\textbf{Recall} & 89.7420 & \textbf{99.2734} & 95.2711 & \textbf{99.6050}
\end{tblr}
\end{table}

\begin{figure*}[t]
    \centering
    \begin{subfigure}[b]{0.45\textwidth}
    \centering
    \includegraphics[width=\textwidth]{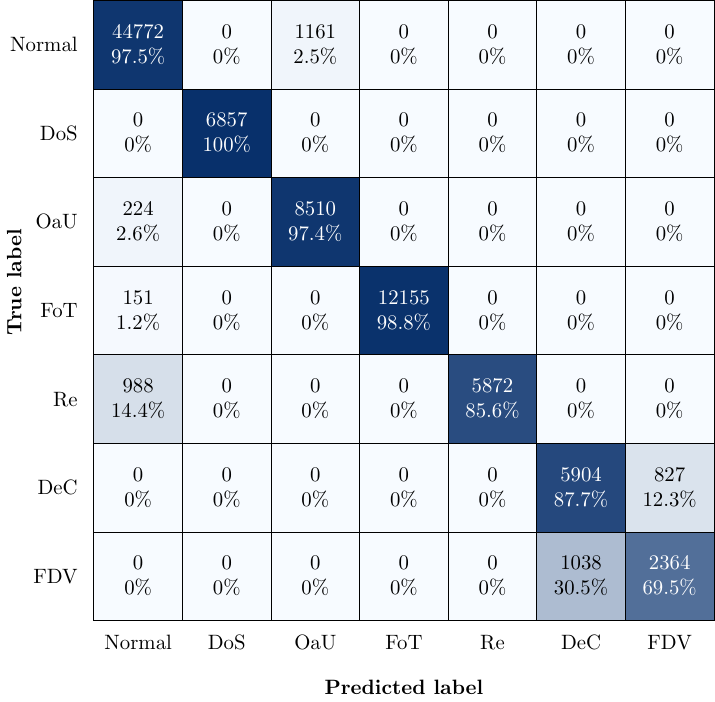}
    \caption{Image transformation preprocessing technique in~\cite{khoa2023collaborative}}
    \label{fig:cfm_compare}
    \end{subfigure}
    \hfill
    \begin{subfigure}[b]{0.45\textwidth}
    \centering
    \includegraphics[width=\linewidth]{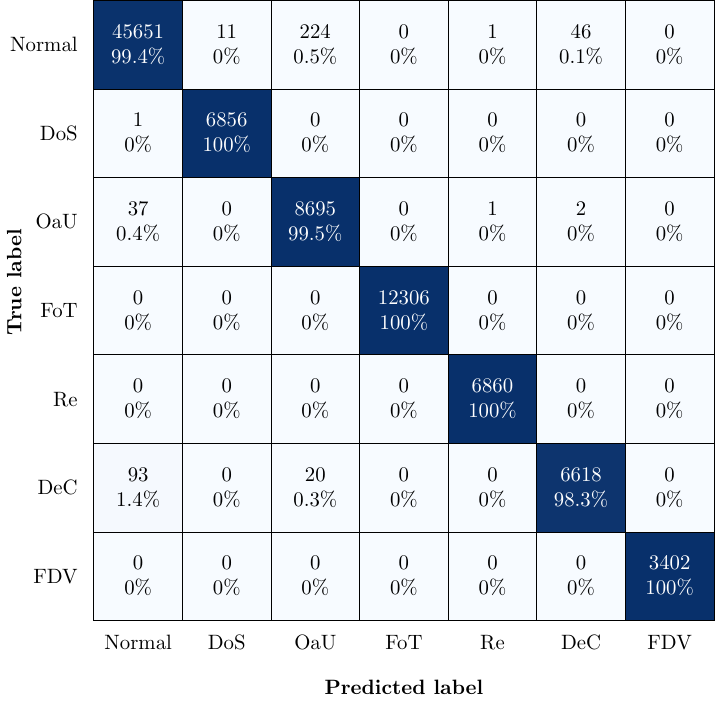}
    \caption{Our proposed NLP-based preprocessing technique}
    \label{fig:cfm_w}
    \end{subfigure}
    \caption{The classification results of ViT-based models with different preprocessing techniques.}
    \label{fig:cfm}
\end{figure*}

\subsubsection{The Impact of Preprocessing Techniques}
In Table~\ref{tab:performance_compare}, we compare the performances of CNN and our ViT-based models using different preprocessing techniques on the BTAT dataset~\cite{khoa2023collaborative}. Using the baseline preprocessing method from~\cite{khoa2023collaborative}, CNN and our ViT-based models achieve accuracies of 93.8\% and 95.16\%, respectively, for detecting attacks in SCs and transactions. In contrast, when applying our proposed NLP-based preprocessing technique, the accuracies significantly improve to 98.8\% for CNN and 99.52\% for ViT-based models. These results demonstrate the effectiveness of our preprocessing method compared to using image transformation alone.

Figure~\ref{fig:cfm}(a) presents the classification results of the ViT-based model using both the baseline and our proposed NLP-based preprocessing methods. The results show that, with the same ViT-based architecture, our approach consistently outperforms the baseline across all attack types. Notably, our method improves the detection accuracy of FDV by approximately 30\%, DeC by 11\%, and Re by 14\% compared to the baseline. The proposed preprocessing technique combines TF-IDF vectorization, concatenated feature vectors, and a unified image combination. This technique helps capture richer semantic and structural information from transaction data, enabling the model to more effectively distinguish among normal and different types of attacks. Due to its improved performance, this preprocessing technique will be used in the following sections to evaluate and compare the performance of our proposed detection framework with existing approaches.

\subsubsection{Model Convergence and Performance Comparison}
\begin{figure}[!htb]
    \centering
    \includegraphics[width=\linewidth]{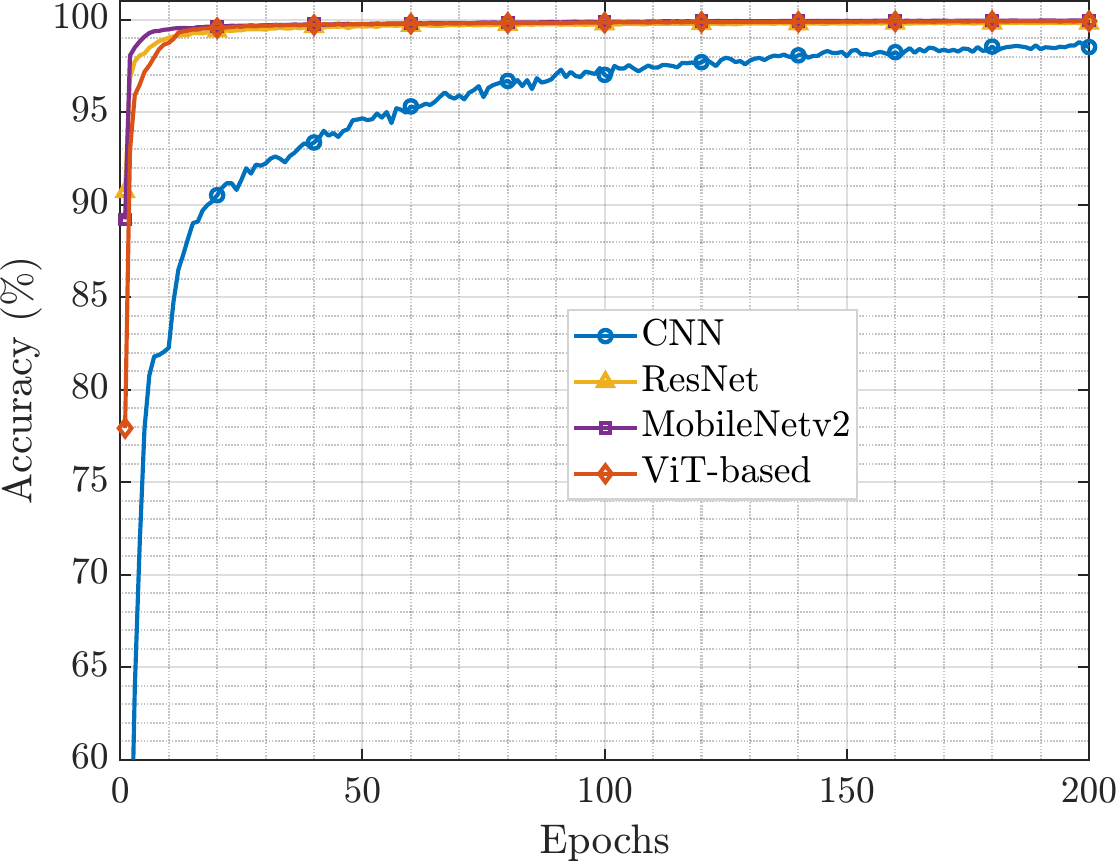}
    \caption{The training convergence of accuracy over different DL models.}
    \label{fig:convergence}
\end{figure}

In this section, we compare the convergence time and performance of our proposed architecture with other relevant models on the training dataset. The comparison contains traditional ML models, including logistic regression (LR), $k$-nearest neighbors (KNN), and support vector machines (SVM)~\cite{Huanhuan2025}, as well as DL models, such as Convolutional Neural Network~(CNN)~\cite{khoa2023collaborative}, ResNet~\cite{Hussain2020}, and MobileNetv2~\cite{Yusuf2023}. As shown in Fig.~\ref{fig:convergence}, the three DL models, ResNet, MobileNetv2, and our proposed ViT-based, require fewer than 20 epochs to achieve near-perfect training accuracy (up to 100\%) during the training phase. On the other hand, the CNN model on the BTAT dataset requires more than 150 epochs to stabilize at approximately 98.5\% accuracy. While our proposed ViT-based model converges almost the same as ResNet and MobileNetv2, it demonstrates a rapid and stable learning process with fewer training epochs compared to the CNN model.
\begin{table*}[!htb]
\centering
\caption{Performance of proposed model versus others.}
\label{tab:perf}
\begin{tblr}{
  width = .8\linewidth,
  colspec = {Q[129]Q[106]Q[106]Q[106]Q[106]Q[120]Q[180]Q[106]},
  column{even} = {c},
  column{3} = {c},
  column{5} = {c},
  column{7} = {c},
  hlines,
  vlines,
}
~ & \textbf{LR}~\cite{Huanhuan2025} & \textbf{KNN}~\cite{Huanhuan2025} & \textbf{SVM}~\cite{Huanhuan2025} & \textbf{CNN}~\cite{khoa2023collaborative} & \textbf{ResNet}~\cite{Hussain2020} & \textbf{MobileNetv2}~\cite{Yusuf2023} & \textbf{ViT-based}\\
\textbf{Accuracy} & 94.8836 & 95.5299 & 96.0198 & 98.8582 & 89.5446 & 99.0080 & \textbf{99.5200}\\
\textbf{Precision} & 94.6416 & 94.2707 & 95.7315 & 98.5740 & 87.9960 & 98.8280 & \textbf{99.4393}\\
\textbf{Recall} & 96.3814 & 94.7914 & 97.6686 & 99.2734 & 93.3549 & 99.2818 & \textbf{99.6050}
\end{tblr}
\end{table*}

\begin{table}
\centering
\caption{Number of training parameters of proposed model versus others.}
\label{tab:perf_params}
\begin{tblr}{
  width = \linewidth,
  colspec = {Q[200]Q[144]Q[192]Q[227]Q[170]},
  column{even} = {c},
  column{3} = {c},
  column{5} = {c},
  hlines,
  vlines,
}
 & \textbf{CNN} & \textbf{ResNet} & \textbf{MobileNetv2} & \textbf{ViT-based}\\
\textbf{No. params} & 806,727 & 11,173,831 & 2,233,543 & 899,559
\end{tblr}
\end{table}

Table~\ref{tab:perf} summarizes the performance of these models on the testing dataset. Overall, our proposed ViT-based model outperforms all other considered approaches, with an accuracy of 99.5\%, a precision of 99.4\%, and a recall of 99.6\%. The conventional ML models, i.e., LR, KNN, and SVM, deliver acceptable results given their simplicity, achieving accuracies of 94.8\%, 95.5\%, and 96\%, respectively. Other DL models, such as CNN and MobileNetv2, produce test results that are consistent with their training performance, achieving accuracies of 98.8\% and 99.0\%, respectively. In contrast, the ResNet model shows a significant drop in testing accuracy, attaining only 89.5\% despite almost perfect results during training. This performance degradation is likely due to overfitting caused by its large number of training parameters relative to the simplicity of the classification task. Furthermore, the detailed classification results of our ViT-based model, as shown in Fig.~\ref{fig:cfm}(b), demonstrate its robustness across all types of attacks in SCs and transactions. It achieves over 99.4\% accuracy in six out of seven classification categories.

\subsubsection{Model Parameter Analysis}
Table~\ref{tab:perf_params} compares the number of trainable parameters across the evaluated DL models. Among the four architectures, ResNet has the largest model size with over 11 million parameters, which explains its tendency to overfit the relatively simple classification task, as observed in the previous evaluation. MobileNetv2, designed for lightweight deployment, has approximately 2.2 million parameters. Our proposed ViT-based model contains only 899,559 parameters, significantly fewer than both ResNet and MobileNetv2, and almost the same as the baseline CNN with 806,727 parameters. Despite its compact size, our proposed ViT-based model achieves the best performance in terms of accuracy, precision, and recall, demonstrating its superior parameter efficiency and suitability for deployment in resource-constrained~environments.

\section{Conclusion}\label{Conclusion}
In this paper, we proposed a novel framework that can efficiently detect cyberattacks in blockchain SCs and transactions in a blockchain system. The framework combined an NLP-based preprocessing technique with a ViT-based model to achieve accurate and efficient detection. In the preprocessing stage, opcode sequences were vectorized using the TF-IDF method and combined with key transaction features, such as gas, input length, and value, to form unified image representations. These images were then classified using the proposed vision-based architecture, which achieved a detection accuracy of 99.5\%, outperforming other baseline models. We also analyzed the number of trainable parameters across different deep learning models to highlight the efficiency of our approach. Our proposed model is lightweight, with a size equal to only 40\% of MobileNetv2 and 8\% of ResNet, making it suitable for deployment on edge devices. In future work, we plan to evaluate the robustness of the framework under adversarial scenarios and extend the approach to support cross-chain transaction analysis.
\balance

\section*{Acknowledgment}
This study was funded by the Australia-Vietnam Strategic Technologies Centre, PTIT, Vietnam

\bibliographystyle{IEEEtran}
\bibliography{library.bib}	
\end{document}